\documentstyle[11pt,newpasp,epsf,twoside]{article}

\begin{document}

\title{Time-series Spectroscopy of EC14026 Stars: Preliminary Results}

\author{Simon J. O'Toole}
\affil{Chatterton Astronomy Department, School of Physics, University of Sydney, NSW 2006, Australia.}
\author{Teresa C. Teixeira}
\affil{Institute of Physics and Astronomy, University of Aarhus, DK-8000 Aarhus c, Denmark}
\author{Timothy R. Bedding}
\affil{Chatterton Astronomy Department, School of Physics, University of Sydney, NSW 2006, Australia.}

\author{Hans Kjeldsen}
\affil{Theoretical Astrophysics Center, University of Aarhus, DK-8000 Aarhus C, Denmark.}

\begin{abstract}
We have obtained time-series spectra for the pulsating hot subdwarf (sdB)
PG1605+072. Previous time-series photometry of this star has shown
maximum amplitude variations of $\sim$0.1 mag and at least 50
periods. The pulsator has the largest amplitude and longest periods of
all the pulsating sdBs (or EC14026 stars) discovered so
far, and appears to be unusual in its class. Preliminary results of a search
for velocity variations are presented here. With these variations, along with
equivalent width measurements, oscillation mode identification should be
possible.
\end{abstract}

\begin{figure}[ht]
\plotfiddle{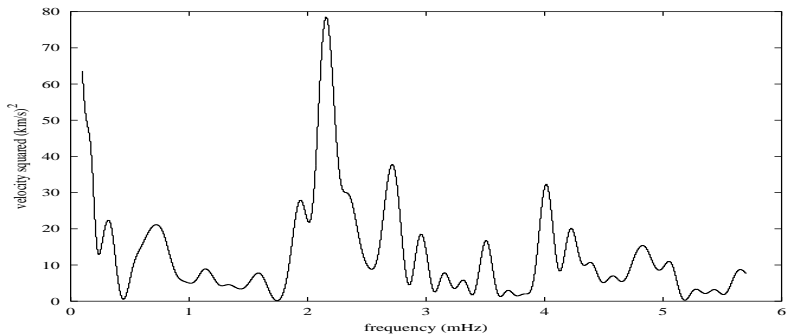}{4cm}{0}{85}{50}{-200}{-20}
\caption{Power spectrum of PG1605+072 using both H$\beta$ and H$\gamma$, averaged over two nights. The peak at $\sim$2.10 mHz corresponds to the strongest photometric amplitude found by Koen et al (1998).}
\end{figure}

\subsection*{Discussion and Further Work}

The main results from our preliminary spectroscopic studies are:

$\bullet$ We have shown the feasibility of using modest size
telescopes (such as the Danish 1.54m) to do time-series spectroscopy
of EC14026 stars.

$\bullet$ Velocity variations in EC14026 stars have been
detected. This confirms that the previous photometric detection of
variability is due to stellar pulsation (not that there were many
doubts).

$\bullet$ We have obtained 11 nights observations on the
Danish 1.54m telescope and as well as 10 nights on the 74 inch telescope at Mt
Stromlo. With these observations we will have the opportunity to use the
complete array of asteroseismological tools to identify oscillation modes, and
hence probe the atmospheres of these stars. Analysis of these data will follow.

\end{document}